\documentstyle[12pt,aaspp4]{article} 
\def\tempest%
{\begin{array}{ccc} 
1 & 1 & 1 \\ 
1 & 1 & 1 \\ 
4 & 3 & 8 
\end{array}}

\def\dls{{d_{\rm ls}}}
\def\dlst{{d_{\rm ls}^2}}
\def\mbr{{M_{\rm brk}}}
\begin{document} 

\title{Measuring the Remnant Mass Function of the Galactic Bulge}

\author 
{Andrew Gould}
\affil{Ohio State University, Department of Astronomy, Columbus, OH 43210} 
\affil{E-mail: gould@astronomy.ohio-state.edu} 
\begin{abstract} 
I show that by observing microlensing events both astrometrically and 
photometrically, the Space Interferometry Mission (SIM)
can measure the mass function
of stellar remnants in the Galactic bulge including white dwarfs, neutron
stars, and black holes.  Neutron stars and black holes can be identified
individually, while white dwarfs are detected statistically from the
sharp peak in their mass function near $M\sim 0.6\,M_\odot$.  
This peak is expected
to be more than twice as high as the ``background'' of main-sequence 
microlenses.   I estimate that of order 20\% of the $\sim 400$ bulge
microlensing events detected to date are due to remnants, but show that these
are completely unrecognizable from their time scale distribution (the only
observable that ``normal'' microlensing observations produce).  
To resolve the white-dwarf peak, the SIM mass measurements must be accurate
to $\sim 5\%$, substantially better than is required to measure the mass
function of the more smoothly distributed main sequence.  Nevertheless, 
SIM could measure the masses of about 20 bulge remnants in 500 hours of 
observing time.

\keywords{astrometry -- black holes --
gravitational lensing -- neutron stars -- white dwarfs} 
\end{abstract} 
\newpage

\section{Introduction} 

	There are two reasons that one would like to measure the mass
function (MF) of stellar remnants in the Galactic bulge, including 
white dwarfs (WDs), neutron stars (NSs), and black holes (BHs).  First,
stellar remnants are one of only two fossils of the era of high-mass
star formation in the bulge (the other being metal abundances).  They
can therefore help determine the stellar MF for masses $M\ga M_\odot$,
a regime that is otherwise virtually inaccessible in the bulge.

	Second, a fair sample of remnants in any environment is difficult
to obtain, so measuring the MF of bulge remnants would shed light on
the study of remnants in clusters, in the field, and perhaps in other
systems as well.  For example, NSs can generally only be detected if they
are pulsars, and it is difficult to estimate what fraction of all NSs are
currently detectable pulsars.  Moreover, the masses of NSs can only
be measured if these are found in suitable binary systems.  While the
range of masses found by this technique is impressively narrow,
$M_{\rm NS} = 1.35\pm 0.04\,M_\odot$ (Thorsett \& Chakrabarty 1999),
this could in principle be due to the narrow range of systems that have
been investigated.  By contrast, WDs can be located in systematic searches
of the solar neighborhood.  However, the WD MF derived from the subsamples
with measurable masses is controversial.  For example, based on a sample
of 164 hot (and so recently formed) WDs, 
Bragaglia, Renzini, \& Bergeron (1995) find a WD MF that is sharply peaked
at $M\sim 0.6\,M_\odot$ with a dispersion (excluding 4 He core WDs)
of $0.16\,M_\odot$ that is mainly generated by a long tail toward high
masses.  On the other hand, Silvestri et al.\ (1999) find a substantially
higher dispersion of $0.25\,M_\odot$ from a sample of 70 cooler, old WDs found
in common proper motion binaries.  They argue that their sample is
more representative of the population as a whole.  To date, BHs have
been found only when they are in relatively close binaries, and even these
are quite difficult to confirm or to measure their masses
(as opposed to obtaining a lower limit).  
The frequency of field BHs is virtually unconstrained.

	The problem of obtaining a remnant MF for the bulge appears
formidable.  A few WDs are bright enough to detect, but typical bulge
WDs are $V\sim 30$.  Serious effort would be required merely to detect
such objects in a high-latitude field, let alone measure their mass.
In crowded bulge fields, optical detection is virtually impossible.  NSs and
BHs are even more difficult to detect.

	However, it is straight forward to detect all three classes of
objects using gravitational microlensing.  Indeed, I will argue below
that of order 80 of $\sim 400$ microlensing events seen toward the bulge
to date (Udalski et al.\ 1994a; Alcock et al.\ 1997) are due to bulge
remnant lenses.  The only problem is that no one has the least idea which
80 they are.

	Here I discuss how observations using
the Space Interferometry Mission (SIM) can
measure important features of the bulge remnant MF.  First, I show that
from their time scales alone (which is normally the only
useful parameter extracted from a microlensing event), the remnants cannot
be separated from the main-sequence (MS) stars, even statistically.  
However, Gould \& Salim (1999)
have shown that if SIM astrometry is combined with photometry from both
the ground and from SIM itself, then SIM can measure of order 100 masses
of microlensing events to $\sim 5\%$ precision in about 500 hours of
observation.  This sample should contain of order 15 WDs and 5 higher-mass
remnants, NSs and BHs. Such a measurement would be adequate to pick out the 
peak of the WD MF and make a rough determination of the frequency of NSs and
BHs.  To make a more detailed measurement of the bulge remnant MF would
require a mission similar to SIM but with a substantially larger aperture.

\section{Main-Sequence Versus Remnant Stars}

	The MF of the bulge MS has been measured in both the optical
(Holtzman et al.\ 1998) and the infrared (Zoccali et al.\ 1999) using
Hubble Space Telescope (HST) observations.  For purposes of this paper,
I will adopt a MS MF that is consistent with those measurements
(but without corrections for binaries),
\begin{equation}
{d N\over d M} = k\biggl({M\over \mbr}\biggr)^\alpha,\qquad
\mbr = 0.7\,M_\odot,
\label{eqn:msmf}
\end{equation}
where $k$ is a constant, and
\begin{equation}
\alpha = -1.3\quad (0.15\,M_\odot<M<\mbr),\qquad
\alpha = -2.0\quad (\mbr<M\la M_\odot).
\label{eqn:powerlaws}
\end{equation}
The upper of limit of $\sim 1\,M_\odot$ is the approximate position of the 
turnoff.  
The lower limit of $0.15\,M_\odot$ comes from the magnitude limit of the
Zoccali et al.\ (1999) observations.  The MF probably continues below
this limit, but at present it is not known how far.  
The change in slope in the
true MF is perhaps less abrupt than is represented in equation
(\ref{eqn:msmf}).  The MF should also be corrected for binaries.  
In Appendix A, I discuss how this correction can be observationally determined.
I will
assume that the slope of $\alpha=-1.3$ observed to $M\sim 0.15$ actually
extends to $M=0.03\,M_\odot$,
\begin{equation}
\alpha = -1.3\quad (0.03\,M_\odot<M<\mbr).
\label{eqn:powerlawrevised}
\end{equation}
It may well extend further, but SIM will
probably not be sensitive to much lower masses because the events are too
short to be alerted in time for SIM observations.  Thus, equations
(\ref{eqn:msmf})--(\ref{eqn:powerlawrevised}) are a rather approximate
representation of the bulge MS MF.
However, I will mainly be using the MS MF for illustration, and for that
purpose, this approximation is quite adequate.   

	I will assume that all MS stars in the range $1\,M_\odot<M<8\,M_\odot$
have now become WDs, and that the total number can be found by extending
the upper MS power law $\alpha=-2.0$ through this higher-mass regime.
That is, $N_{\rm WD}= (7/8) k \mbr$.  Of course, there is no evidence
whatever that the slope does continue in this regime.  A more popular slope
is the Salpeter value $\alpha=-2.35$.  Had I chosen this steeper slope, 
my estimate
for $N_{\rm WD}$ would be reduced by a factor 0.80.  For the distribution
of WD masses, I adopt the MF shown in Figure 11c of Bergeron et al.\ (1995)
based on observations of 164 hot WDs.  I assume that all MS stars 
$8\,M_\odot<M<40\,M_\odot$ become NSs, with masses that are centered
at $M=1.35\,M_\odot$ and with Gaussian dispersion of $0.04\,M_\odot$
(Thorsett \& Chakrabarty 1999). I assume that all MS stars 
$40\,M_\odot<M<100\,M_\odot$ become BHs, with masses that are centered
at $M=5\,M_\odot$ and with Gaussian dispersion $1\,M_\odot$.  I have no
justification for this BH distribution, but since BHs with $M\ga 2\,M_\odot$
will be easily identified by SIM whatever their mass (see \S 3), 
the particular form of their distribution can be
fixed arbitrarily.  I also assume that the power law $\alpha=-2$ extends
throughout this entire regime.  Again, this assumption is arbitrary but it
is appropriate for purposes of illustration.

	With these assumptions, the fractions of numbers $N$ of
objects in the four classes,
MS stars (including brown dwarfs $M>0.03\,M_\odot$), WDs, NSs, and BHs are
(89:10:1:0.2).  The fractions of the total mass (which scales $\propto N M$)
are (69:22:6:3).  The fractions of microlensing events (which scale 
$\propto N M^{1/2}$) are
(79:17:3:0.8).  That is, of order 20\% of the bulge microlensing events
are due to remnants.  

\subsection{Combined Mass Function}

	Figure \ref{fig:one}a shows the distribution of microlensing
events per unit log mass, as a function of log mass and classified by type
of object.  It is normalized to a total of 100 bulge microlensing events.
Gould \& Salim (1999) showed that masses of approximately 100 microlensing
events could be measured by SIM.   Of these, only $\sim 80\%$ should be due
to bulge (rather than disk) lenses.  Nevertheless,
for simplicity I normalize to 100 events.
The WD bins in Figure \ref{fig:one}a are $0.05\,M_\odot$ wide.
Notice that the peak WD bin (which contains about 1/3 of
all the WDs) stands well above the ``background'' of MS stars, and that
there are several adjacent WD bins that are comparable to the MS
background.  If 100 individual masses were measured 
then about 6 stars would be in this peak bin and another
6 would be in the two neighboring bins toward higher mass.  From these
three bins alone, I find a total signal-to-noise ratio,
S/N $=[\sum_i N^2_{{\rm WD},i}/( N_{{\rm WD},i} + N_{{\rm MS},i})]^{1/2}
\sim 3$,
where $N_{{\rm WD},i}$ and  $N_{{\rm MS},i}$ are the numbers of WD and MS
stars in bin $i$.  This implies
a statistical accuracy in the WD frequency of about 33\%.  The NSs and
BHs would be free of any MS background, so their frequency measurement
would be limited by Poisson errors, roughly 50\% for a combined total
of 4 stars.

\subsection{Time Scale Distribution at Fixed Mass}

	I have previously shown that if one assumes that the bulge
lenses are distributed as $r^{-2}$, that the lenses and sources
each have Gaussian velocity distributions with one-dimensional dispersion
$\sigma$, and that all the sources are at the Galactocentric distance,
then the time scale distribution for lenses at fixed mass $M$ is given by
\begin{equation}
{d\Gamma\over d \ln t_{\rm E}} 
\propto t_{\rm E}^{-1}
\int_0^\infty d z{z^2\over z^2 +(t_{bM}/t_{\rm E})^4}\exp(-z)
\label{eqn:dgammadlnt}
\end{equation}
where
\begin{equation}
t_{bM} \equiv {(G M b)^{1/2}\over \sigma c},
\label{eqn:tbdef}
\end{equation}
and $b$ is the impact parameter of the line of sight relative to the
Galactic center (Gould 1995b).  My excuse for assuming that all the 
sources were at the Galactocentric distance was that it made the problem
``analytically tractable'', but probably I just forgot to put on my
thinking cap:  If all the sources were assumed to be at the Galactocentric
distance, then the distribution of source-lens separations, $\dls$
(not weighted by the lensing cross sections) would be just 
$f_0(\dls;b)=K(b^2 + \dlst)^{-1}$, where $K$ is a constant.
If we now assume that the lenses and sources are both distributed as
$r^{-2}$, then the distribution of separations is
\begin{equation}
f(\dls;b) = {\int d x\,d y\, f_0(x;b)f_0(y;b)\delta(y-x-\dls)\over
\int d x\, f_0(x;b)} = {2 K\over \dlst + 4 b^2} = 2 f_0(\dls;2b),
\label{eqn:fdls}
\end{equation}
where $\delta$ is a Dirac $\delta$-function.
That is, equation (\ref{eqn:dgammadlnt}) remains valid but with
\begin{equation}
t_{bM} \rightarrow {(2 G M b)^{1/2}\over \sigma c}.
\label{eqn:tbrevised}
\end{equation}

	Figure \ref{fig:one}b shows the distribution of events per log
time scale as a function of $\log (t_{\rm E}/t_{bM})^2$. The reason for
using this scale (which of course is simply double the natural scale
of $\log[t_{\rm E}/t_{bM}]$) is that for microlensing events 
$t_{\rm E}^2\propto M$.  Thus Figures \ref{fig:one}a and
\ref{fig:one}b can be directly compared.  This comparison reveals that
the width of the time scale distribution at fixed mass is enormously larger 
than the scale on which the remnant populations have structure and is
therefore likely to obliterate any distinctive traces of the remnants.

	Figure \ref{fig:one}c demonstrates that this is indeed the case.
It shows the distribution of time scales that will be observed from
the mass distribution in Figure \ref{fig:one}a given the time scale 
distribution
at fixed mass shown in Figure \ref{fig:one}b. Figure \ref{fig:one}c is simply
the convolution of Figures \ref{fig:one}a and \ref{fig:one}b.
The time scale normalization parameter is 
$t_{bM_\odot}=(2 G M_\odot b)^{1/2}/\sigma c$.   The WD peak is now so
spread out that it cannot be distinguished from the MS.  The NS and BH peaks
are also spread out, but the main problem is that, in constrast to
Figure \ref{fig:one}a, they are now overwhelmed by background from the
much more numerous MS stars whose time scale distribution is smeared out
over the entire range illustrated in the figure.  The conclusion is that
while a significant fraction of microlensing events are due to remnants,
these remnants will never be recognized as such from their observed
time scales alone.

\section{Measuring the Remnant Mass Function with SIM}

	For bright bulge sources ($I\sim 15$) SIM can measure the masses
of bulge lenses with $\sim 5\%$ precision in about 5 hours of observing time
(Gould \& Salim 1999).
 From Figure \ref{fig:one}a (where the WDs are binned by $0.05\,M_\odot$), 
it is clear that approximately this precision is
required to resolve the WD peak and so be able to efficiently separate
the WDs from the MS background.  Indeed, the mass errors reported by
Bragaglia et al.\ (1995) when they constructed the WD MF used as a basis
for Figure \ref{fig:one}a are of order 5\%.  Thus, if the bulge WD MF is
similar to the hot WD MF investigated by Bragaglia et al.\ (1995),
then the distribution of masses measured by SIM (including 5\% errors) should
look fairly similar to the one shown in  Figure \ref{fig:one}a.

	Note that if one were concerned only to measure the mass function
of main-sequence stars, the precision requirements could be relaxed
significantly because the main sequence does not exhibit structure on
small scales.  In fact, for the main sequence
the precision is required more to measure the
lens-source relative parallax (which has the same fractional error as
the mass -- Gould \& Salim 1999) in order to be able to separate bulge
lenses from disk lenses.

	The measurement requirements for NSs and especially BHs are less 
severe than for WDs.  
NSs are $\sim 35\%$ heavier than turnoff stars, so if their masses
could be measured to $\sim 10\%$, they could be reliably distinguished
from MS stars.  Moreover, within the model I am considering,
events due to upper MS stars ($0.7\, M_\odot< M < 1.0\,M_\odot$) are only
about three times more common than those due to NSs, so the tail of the
MS event distribution will not seriously corrupt the measurement of the
NS frequency.  

	Unfortunately, for fixed SIM observation time, the fractional
error in the mass measurement grows with mass (Gould \& Salim 1999).  The
 reason for this is that two quantities must be separately measured
to determine $M$, the angular Einstein radius $\theta_{\rm E}$ 
(Boden, Shao, \& Van Buren 1998) and the size of the Einstein radius 
projected onto the plane of the observer, $\tilde r_{\rm E}$ (Gould 1995a).  
Specifically $M= (c^2/4G)\tilde r_{\rm E}\theta_{\rm E}$.
Both $\theta_{\rm E}$ and $\tilde r_{\rm E}$ scale  $\propto M^{1/2}$.
As $\theta_{\rm E}$ grows, the astrometric deviation grows with it and
so becomes easier to measure. However, $\tilde r_{\rm E}$ is measured
from the difference in the photometric light curves as seen from the
Earth and SIM.  The bigger $\tilde r_{\rm E}$, the closer the Earth and
SIM are in the projected Einstein ring, and the harder it is to measure
the difference in the light curves as seen from the two observatories.
Thus, observations that would be sufficient to obtain 5\% errors for 
``typical'' stars $(M\sim 0.3\,M_\odot)$, achieve only about 10\% precision for
NSs of $1.35\,M_\odot$, and 30\% precision for BHs of $5\,M_\odot$.
However,
since BHs are separated from the MS by a factor of a few in mass, even errors
of order 30\% would be sufficient to recognize them as such.  

	Thus, the program outlined by Gould \& Salim (1999) would be
adequate to recognize NSs and BHs individually whenever the events were
observed, and would also suffice to recognize the peak of the WD mass
function.  The main limitation of this program is its modest statistics:
fewer than 1 BH event and only 3 NS events are expected.  

	In principle it would be possible to overcome this problem
simply by observing more events.  However, the observation time scales
inversely as the source flux.  There are a limited number of events
with bright sources.  If one wanted to increase the number of measurements
by a factor of two and maintain the same errors, one would be forced to
observe fainter sources and hence substantially increase the observing
time per source.  Thus, the observation time would grow much more
rapidly than the number of measurements.  It therefore appears that 
the only feasible route
to fainter sources (and thus substantially more events) is a mission
similar to SIM but with apertures that are larger than the 25 cm mirrors
on SIM.

{\bf Acknowledgements}:
I would like to thank S.\ Salim for calculating the SIM errors for NS
and BH masses and B.S. Gaudi for a careful reading of the manuscript.
This research was supported in part by grant AST 97-27520 from the NSF
and in part by grant NAG5-3111 from NASA.

\appendix
\renewcommand{\theequation}{\thesection\arabic{equation}}
\section{How To Determine the Bulge Binary Distribution}

      The correction for binaries remains an important uncertainty in the
mass budget of the bulge.  Binaries in the bulge, in sharp contrast to those
in the  solar neighborhood (e.g. Duquennoy \& Mayor 1992) or even in globular 
clusters  (Hut et al.\ 1992), are virtually unstudied.  Here, I briefly 
outline how they could be.  Close binaries of roughly solar-mass primaries
could be studied from eclipsing binaries found in microlensing studies 
(Udalski et al.\ 1994b, 1995a,b) and radial-velocity measurement 
of clump giants.
The probability to be an eclipsing system declines inversely as the semi-major
axis, and the fraction of the period spent in eclipse falls at the same rate,
so this technique can probably be extended only to $\sim 100$ stellar radii,
or about 1 AU for turnoff sources.  For substantially
brighter sources the eclipses are
too shallow to easily detect and for substantially fainter sources the
source is too faint to monitor with current or foreseeable programs.
An extensive radial velocity survey requires bright sources of which the
most numerous are clump giants.  These can be searched for companions from
$\sim 0.5\,$AU (inside of which the companion may have been affected during
the red giant phase) to $\sim 5\,$AU (beyond which the orbital periods become
too long to monitor).  

      Microlensing surveys can effectively
search for binaries over the range $0.2\la b \la 30$ where 
$b\equiv r_p/r_{\rm E}$, $r_p$ is the projected separation, and
$r_{\rm E}$ is the Einstein radius which for bulge lensing events is roughly
given by $r_{\rm E}\sim 3.5\,(M/M_\odot)^{1/2}\,\rm AU$.  For 
$b\la 0.2$, the binary microlensing event behaves photometrically like
a point mass (Gaudi \& Gould 1997) and so cannot be recognized
(although binaries with even smaller 
separations can be recognized astrometrically,
Chang \& Han 1999).  For
$0.2\la b\la 2$ and for a significant fraction of events, 
the binary produces characteristic caustic
structures that are easily recognized.  For $2\la b\la 30$, the
binary gives rise to two-peaked events (Di Stefano \& Scalzo 1999).  
Although some of the automated
routines used by microlensing search teams might throw these events out
on the grounds that ``microlensing events do not repeat'', there are
hundreds of events that have been recognized in real time (``alerts'')
and these could be searched for second peaks up to several years after the
first bump.  The major limitation is that the probability of a second bump
falls off as $b^{-1}$.  In addition, one must wait
$b$ Einstein crossing times (each typically 
$t_{\rm E}\sim 30\,(M/M_\odot)^{1/2}\,{\rm days}$) for the bump to occur,
and this may exceed the duration of the microlensing experiment which in
general have long, but finite, lifetimes.

    Finally, it should be possible to search for common proper motion
pairs by comparing two Hubble Space Telescope observations.  For example
the first epoch could be taken from Holtzman et al.\ (1998).  It should
be possible to measure positions from a single image accurate to
2 mas (I.\ King 1999, private communication), and so with a five year
baseline (and multiple exposures at each epoch) easily distinguish
common proper motion pairs from optical binaries whose typical random
proper motions are 3~mas~yr$^{-1}$.  The limitation is that even for the
HST Planetary Camera, the projected
separations must be $0''\hskip-2pt .15$ or more
(see Table 1 of Gould et al.\ 1995) depending on the magnitude difference
just to detect separate stars.  This corresponds to $r_p\sim 1200\,$AU at the
Galactocentric distance.

    By combining all these techniques, one could cover separations
$r_p\la 100\,$AU and $r_p\ga 1200\,$AU.  
Although a substantial range of separations
would be inaccessible, one could determine whether the basic pattern
found by Duquennoy \& Mayor (1991) for solar-type primaries in the solar
neighborhood also holds true for the bulge.  If it did, one could interpolate
into the unobserved interval $100\,{\rm AU}\la r_p\la 1200\,{\rm AU}$ using
the local data.

\newpage

\begin{figure}
\caption[junk]{\label{fig:one}
Rates of microlensing events toward the bulge by mass (panel {\it a}) and 
time scale (panel {\it c}) for
MS stars and brown dwarfs ($0.03\,M_\odot<M<1\,M_\odot$) 
({\it bold dashed curve}) 
and WD, NS, and BH remnants ({\it solid curves}).  The
total is shown by a {\it bold solid curve}.   The mass model ({\it a}) 
is described in \S\ 2.  In particular, the WD distribution is shown
in $0.05\,M_\odot$ bins taken from  Bragaglia et al.\ (1995).
The mass model is convolved with the time scale distribution 
at fixed mass ({\it b}) derived in \S\ 2.2, 
to produce the observable time scale distribution ({\it c}).  The abcissas
of panels ({\it b}) and ({\it c}) contain $\log t_{\rm E}^2$ rather
than $\log t_{\rm E}$ so that they can be directly compared with panel
({\it a}), since $ t_{\rm E}^2\propto M$.  All three classes of remnants
are clearly identifiable in the mass distribution, but are utterly lost
in the time scale distribution.  The normalizations in panels
({\it a}) and ({\it c}) are for 100 events.  Panel ({\it b}) is
normalized to unity.
}
\end{figure}

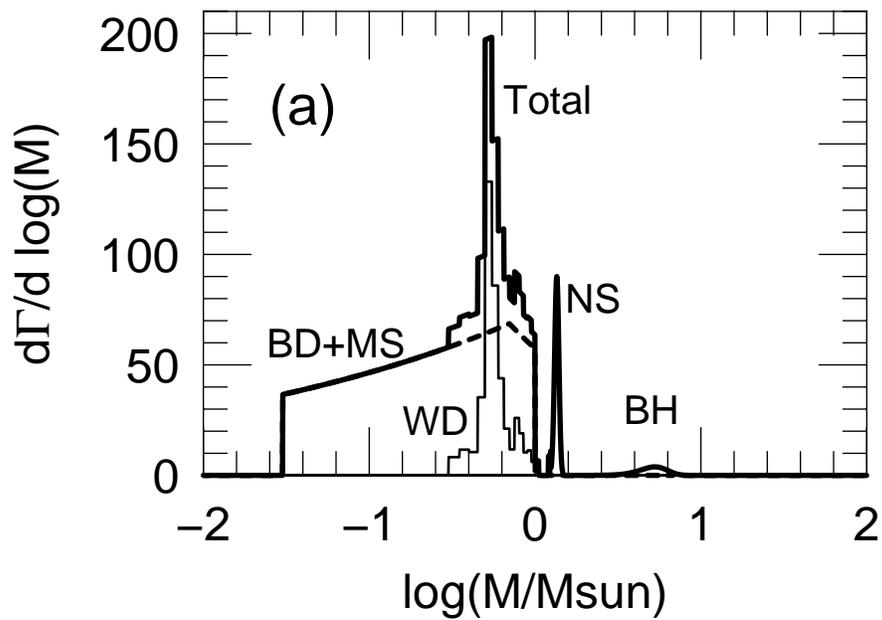
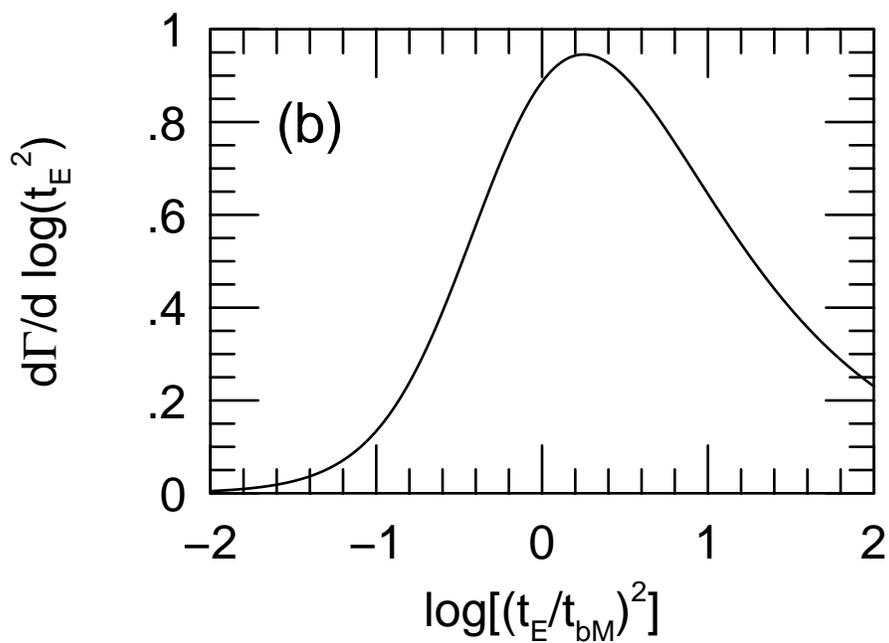
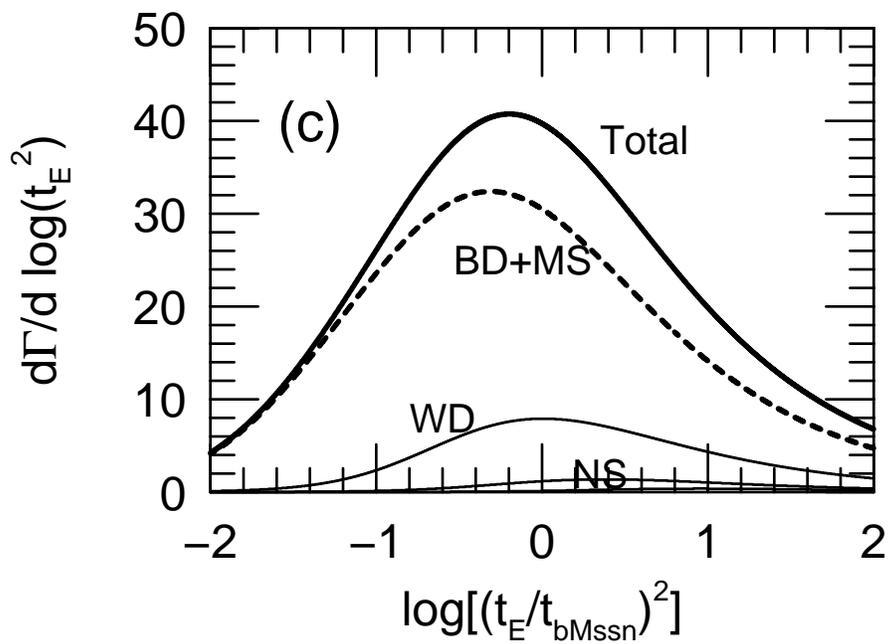
\end{document}